\documentstyle[12pt,fleqn]{article}
\def\soc{{\rm C}_{60}}
\def\rug{{\rm C}_{70}}

\def\la{\langle}
\def\ra{\rangle}
\def\beeq{\begin{equation}}
\def\eneq{\end{equation}}
\def\beeqa{\begin{eqnarray}}
\def\eneqa{\end{eqnarray}}

\addtocounter{section}{-1}
\begin{document}

\begin{center}

\vspace{2in}

{\large {\bf{A strongly-interacting model for antiferromagnetism\\
in one-dimensional C$_{\bf 60}$-polymers:
{\mbox{\boldmath $A$}}$_{\bf 1}$C$_{\bf 60}$
} } }

\vspace{1cm}

{\rm Kikuo Harigaya\footnote[1]{E-mail address:
harigaya@etl.go.jp; URL: http://www.etl.go.jp/People/harigaya/}}\\

\vspace{1cm}

{\sl Fundamental Physics Section,\\
Electrotechnical Laboratory,\\
Umezono 1-1-4, Tsukuba, Ibaraki 305, Japan}

\vspace{1cm}

(Received~~~~~~~~~~~~~~~~~~~~~~~~~~~~~~~~~~~)
\end{center}

\vspace{1cm}

\noindent
{\bf ABSTRACT}\\
We study the antiferromagnetic phase of the $\soc$-polymer doped
with one electron per one $\soc$ by using a tight-binding model
with long-range Coulomb interactions.  The model is solved
by the unrestricted Hartree-Fock approximation, and a phase
diagram is obtained by changing Coulomb interaction strengths
and a conjugation parameter between $\soc$ molecules.
We find that the antiferromagnetism is well described by the
present model.  The comparison with the photoemission studies
shows that the new band around the Fermi energy of the $A_1$$\soc$
phase can be explaind by the extremely large intrusion of an
energy level into the gap of the neutral system.  This indicates
that the interaction effects among electrons are important in doped
$\soc$-polymers.

\mbox{}


\pagebreak

Since the discovery of the linear (one-dimensional) $\soc$-polymers
in alkali-metal doped $\soc$ crystals: $A_1\soc$ ($A=$K, Rb, Cs),$^{1-4}$
their solid state properties have been intensively investigated.
One electron per one $\soc$ is donated in the polymer chain.
It seems that Fermi surfaces exist in high temperatures, but the
system shows antiferromagnetic correlations in temperatures lower
than about 50 K.$^1$  Muon-spin-relaxation experiments$^5$ have
shown the presence of the local magnetic field, suggesting a static
magnetic ordering.  In the $\soc$-polymer, $\soc$ molecules are
arrayed in a linear chain.  The bonds between $\soc$ are formed by
the [2+2] cycloaddition mechanism.  The lattice structure is shown
in Fig. 1.  The sites, 1-4, are the lattice points of the
four-membered ring.

Several calculations of the electronic structures have been
reported.  One-dimensional as well as three-dimensional model
systems have been studied theoretically.  The one-dimensional
calculations in the literatures are as follows:\\
(1)  A tight-binding calculation of a linear chain$^6$ has
been performed, and the relation to the antiferromagnetic
ground state has been discussed with the assumption that the
system is simulated well by the one-dimensional Hubbard model
with the effective inter-$\soc$ hopping interaction $t_{\rm eff}$
and the on-ball Couloumb repulsion $U_{\rm eff}$.\\
(2)  A semiempirical tight-binding model$^7$ analogous to the
Su-Schrieffer-Heeger (SSH) model$^8$ of conjugated polymers
has been proposed, and the possibility of the charge density
wave state has been pointed out.\\
(3)  We have applied another version of the SSH-type model
to the $\soc$-polymer chain.$^9$  We have discussed a possible
metallic state of the neutral $\soc$-polymer in the limit of
the large intermolecular interactions.  (Recently, this model
has been extended to the two-dimensional $\soc$-polymers.$^{10}$)\\
The band calculation for three-dimensional systems has also
been done, for example:\\
(4)  A tight-binding calculation for Rb$_1$$\soc$ has been
performed with using the crystal structure obtained from the X-ray
diffraction.$^{11}$  It has been discussed that the nesting
property of the Fermi surfaces remains even if the interchain
interactions give rise to a certain dispersion in the directions
perpendicular to the polymer chains.

The purpose of this paper is to propose a microscopic model
for the antiferromagnetism in $\soc$-polymers and to show that
the antiferromagnetic ground state can be actually explained
by our model with strong electron-electron interactions.
Our model is one dimensional and has two contributions.
The first term is the hopping interactions along the polymer
backbone structure which has been used in Ref. 9.  The second term
is the Coulomb interaction among electrons.  We use the long-range
Ohno potential which we have used in the previous paper.$^{12}$
In Ref. 6, the relevance of the Hubbard model, where the $\soc$
molecule is regarded as a site, has been discussed.  We will
compare the present calculation with the assumption of the
Hubbard model.  Finally, we discuss the relevance of the model
with strong electron-electron interactions comparing with the
recent photoemission experiments.$^{13}$.  We will show that the
large shift and intrusion of the LUMO band of the neutral system
are the results of the strong interactions among electrons rather
than the polaron effects due to the Jahn-Teller interactions.$^{14}$

We use a tight-binding model with Coulomb interactions
among electrons, in order to discuss the antiferromagnetism in
$\soc$-polymers.  The model is the following:
\beeqa
H &=& H_{\rm pol} + H_{\rm int}, \\
H_{\rm pol} &=&  - at \sum_{l,\sigma}
\sum_{\langle i,j \rangle = \langle 1,3 \rangle,\langle 2,4 \rangle}
( c_{l,i,\sigma}^\dagger c_{l+1,j,\sigma} + {\rm H.c.} ) \nonumber \\
&-&  (1-a)t \sum_{l,\sigma}
\sum_{\langle i,j \rangle = \langle 1,2 \rangle,\langle 3,4 \rangle}
( c_{l,i,\sigma}^\dagger c_{l,j,\sigma} + {\rm H.c.} )  \nonumber \\
&-& t \sum_{l,\sigma} \sum_{\langle i,j \rangle = {\rm others}}
( c_{l,i,\sigma}^\dagger c_{l,j,\sigma} + {\rm H.c.} ), \\
H_{\rm int} &=& U \sum_{l,i}
(c_{l,i,\uparrow}^\dagger c_{l,i,\uparrow} - \frac{n_{\rm el}}{2})
(c_{l,i,\downarrow}^\dagger c_{l,i,\downarrow}
- \frac{n_{\rm el}}{2}) \nonumber \\
&+& \sum_{l,l',i,j} W(r_{l,l',i,j})
(\sum_\sigma c_{l,i,\sigma}^\dagger c_{l,i,\sigma} - n_{\rm el})
(\sum_\tau c_{l',j,\tau}^\dagger c_{l',j,\tau} - n_{\rm el}).
\eneqa
In Eq. (1), the first term is the tight binding part of the
$\soc$-polymer, and the second term is the Coulomb interaction
potential among electrons.  In Eq. (2), $t$ is the hopping integral
between the nearest neighbor carbon atoms; $l$ means the $l$th
$\soc$ molecule and $\langle i,j \rangle$ indicates the pair of the
neighboring $i$th and $j$th atoms; the atoms with $i=1 - 4$ of the
four-membered ring are shown by numbers in Fig. 1 and the other $i$
within $5 \leq i \leq 60$ labels the remaining atoms in the same
molecule; $c_{l,i,\sigma}$ is an annihilation operator of the electron
at the $i$th site of the $l$th molecule with spin $\sigma$;
and the sum is taken over the pairs of neighboring atoms.
Eq. (3) is the Coulomb interactions among electrons.
Here, $n_{\rm el}$ is the number of electrons per site;
$r_{l,l',i,j}$ is the distance between the $i$th site of the $l$th
$\soc$ and $j$th site of the $l'$th $\soc$; and
\beeq
W(r) = \frac{1}{\sqrt{(1/U)^2 + (r/r_0 V)^2}}
\eneq
is the Ohno potential used in Ref. 12.  The quantity $W(0) = U$ is
the strength of the onsite interaction; $V$ means the strength of
the long range part; and $r_0 = 1.433$\AA~ is the mean bond length
of the single $\soc$ molecule.  We use the long-range interaction
in order to discuss the antifferomagnetism which extends over
many molecules.  The onsite Hubbard interaction seems so short
in view of the dimension of the long-range order.

The parameter $a$ controls the strength of conjugations between
neighboring molecules.  This parameter has been introduced in the
previous papers.$^{9,10}$  When $a=1$, the $\sigma$-bondings between
atoms, 1 and 2, 3 and 4, in Fig. 1 are completely broken and the
orbitals would become like $\pi$-orbitals.  The bonds between the
atoms, 1 and 3, 2 and 4, become double bonds.
As $a$ becomes smaller, the conjugation between the neighboring
molecules decreases, and the $\soc$ molecules become mutually
independent.  In other words, the interactions between molecules
become smaller in the intermediate $a$ region.  In this case,
the operator $c_{l,i,\sigma}$ at the lattice sites of the
four-membered rings is one of the relevant linear combinations of
the $sp^3$ orbitals.  Here, we assume a possibility of local
conjugations at the four-membered rings.  When $a=0$,
the $\soc$ molecules are completely disconnected each other.

The present unit cell consists of two $\soc$ molecules, because
the antiferromagnetism is the alternation of the up and down spins
along the chain direction.  We consider the electron doped case, so the
electron number is 61 for one $\soc$.  We perform the unrestricted
Hartree-Fock approximation for operators of electrons.  The
Hartree-Fock order parameters are determined by the iteration method
using the two-molecule unit cell in the real space.  The dispersions
of the energy bands are determined by the Bloch theorem.
We will change the parameter, $a$, within $0 \leq a \leq 0.5$.
This might be appropriate for the cases of the weak inter-$\soc$
conjugations for $\soc$-polymers under room pressures.
For Coulomb parameters, we put $V = U/2$ from our experiences
of the calculations on optical excitations in $\soc$ and $\rug$.$^{12}$
The value of $U$ is changed within $0 \leq U \leq 3.0t$.
Most of the data with the energy dimension are shown in units of $t$.

Figure 1 shows the unit cell of the $\soc$-polymer with the
bond lengths which are used for the present calculations.
The length 9.138\AA~ between the centers of $\soc$ and bond
lengths, 1.44\AA, 1.90\AA, and 1.51\AA, around the four-membered
ring are taken from the crystal structure data.$^2$  The
other bond lengths, 1.40\AA~ (for the short bonds) and
1.45\AA~ (for the long bonds), are the same as used in the
previous paper.$^{12}$

The magnitude of the ordered spin per one $\soc$ is calculated,
and is shown in Fig. 2 changing $a$ and $U$.  The magnitude
is finite when the magnetic order appears.  As $U$ becomes stronger,
the magnetization appears and becomes larger.  This is a natural
consequence.  When $a$ is smaller, the itinerancy of electrons
in the chain direction decreases, and thus the antiferromagnetism
appears more easily.  This feature is realized in the present
calculation: the critical value of $U$, where a finite magnetization
begins to appear, decreases as $a$ becomes smaller.  The larger
bond length of $\langle 1,3 \rangle$ than that of $\langle 1,2
\rangle$ would be consistent with the assumption $0 \leq a \leq 0.5$,
but we have used the bond lengths shown in Fig. 1.  Even though
different bond lengths of $\langle 1,2 \rangle$ and $\langle
1,3 \rangle$ are assumed, the small $a$ is relevant in the present
phenomenological model.

The calculated results are summarized as a phase diagram in Fig. 3.
The metallic phase region is named as M.  And the antiferromagnetic
phase is named as AF.  The antiferromagnetism appears in the small
$a$ region when $U$ is taken constant.  The phase boundary between two
phases is an almost linear line.  In Ref. 6, the applicability of the
one-dimensional Hubbard model is discussed.  When this discussion
is adopted to the present model Eq. (1), the model can be mapped
to the following Hamiltonian:
\beeq
H = - t_{\rm eff} \sum_{\la l,l' \ra,\sigma}
(a^\dagger_{l,\sigma} a_{l',\sigma} + {\rm H.c.})
+ U_{\rm eff} \sum_{l} n_{l,\uparrow} n_{l,\downarrow},
\eneq
where $t_{\rm eff}$ is an effective hopping integral between
neighboring $\soc$ molecules; $U_{\rm eff}$ is the on-ball
Coulomb repulsion; $a_{l,\sigma}$ is an annihilation operator
of an effective orbital of the doped electron at the $l$th molecule;
and $n_{l,\sigma} = a_{l,\sigma}^\dagger a_{l,\sigma}$.
In the one-dimensional Hubbard model, it is known that the
antiferromagnetism appears even when $U_{\rm eff}$ is small but
positive.   But, there is a parameter region where a magnetic
order is not present in Fig. 3.  Thus, we find that the
mapping onto the Hubbard model is not always relevant to
all the parameter set.  The mapping could be used for the
case with small $a$ and large $U$.
In Ref. 6, the antiferromagnetism has been discussed in
connection with the polaron in the doped $\soc$.$^{14}$  In the
present paper, electron-phonon interactions, whose effects on the
band structures are only small, are not considered, and only
electron-electron interactions are taken into account.
Therefore, the antiferromagnetism is not directly related with the
polaron formation due to the Jahn-Teller effects.$^{14}$  Rather,
we could regard the antiferromagnetism as one of the spin
density wave states in the itinerant electron systems.
It is of course that the doped extra electrons tend to
have large amplitudes at certain sites on the surface of
the $\soc$ molecules.  However, the main origin of the
localization would be the strong electron-electron interactions
rather than the Jahn-Teller effects.

It is useful to look at how the band structures change
upon the formation of the antiferromagnetic order.  The
dispersions of the bands of the system before the electron
doping and without the magnetic order are shown for
$(a,U,V) = (0.3, 2.4t, 1.2t)$ in Fig. 4.  Here, the electron
number is 60 per one $\soc$, and the unit cell consists
of one $\soc$.  The band structures of the antiferromagnetic
system are shown for the same $(a,U,V)$ in Fig. 5.
The unit cell becomes doubled, so the first Brilloune zone
is half of that of Fig. 4.  The length of the abscissa
of Fig. 5 is expanded by 2 from that of Fig. 4 in order
to display the band structures clearly.  The energy bands in
Fig. 5 become narrower.  The band named ``HOMO" in Fig. 4 changes into the
HOMO-1 band and HOMO-2 band in Fig. 5.  There is a small energy gap
between the HOMO-1 and HOMO-2 bands at the boundary of the first
Brilloune zone in Fig. 5.  In the same way, the band named
``LUMO" of Fig. 4 is the origins of the ``HOMO" band and ``LUMO" band
of Fig. 5.  There is a large energy gap at the zone boundary
as well.  These energy gaps are opened by the formation
of the antiferromagnetism.

It should be noted that the energy band intrudes into the original
HOMO-LUMO band gap of the neutral system extremely largely.  The HOMO
band of the system with the antiferromagnetism even locates near the
center of the original HOMO-LUMO band gap.  This intrusion is much
larger than that in the polaron formation due to the Jahn-Teller
interactions in the doped $\soc$,$^{14}$ and cannot be obtained from
models with electron-phonon interactions only.  Actually, the new band
has been observed in the photoemission studies of the orthorhombic
phase of Rb$_1$$\soc$.$^{13}$  Such the large changes in band
structures might be the evidence of strong electron-electron
interactions in the $\soc$-polymers rather than the results of
the Jahn-Teller distortion due to the electron-phonon interactions.
The positons of the HOMO band and HOMO-1 band of Fig. 4 do not shift
so much upon the formation of the antiferromagnetism.
This is also consistent with the photoemission studies.$^{13}$

In this paper, we have not considered possible interchain
interactions which might effect on the dispersion of bands
perpendicular to the polymer direction, because they might not
change the one-dimensional band structures so much,$^{11}$
and because the low dimensionality of the polymer-chain
might be favorable for the orderings to occur due to
interactions among itinerant electrons.

To summarize, we have studied the antiferromagnetic phase of
the $\soc$-polymers
by microscopic calculations.  A tight-binding model with long-range
Coulomb interactions has been proposed.  The model has been solved
by the unrestricted Hartree-Fock approximation, and the antiferromagnetic
phase boundary has been obtained by changing the Coulomb interaction
strengths and the conjugation parameter between $\soc$ molecules.
We have found that the antiferromagnetism is well described by the
present model.  We have discussed that the formation of the magnetic
order is closely related with the strong interactions rather than
the Jahn-Teller mechanisms.  Finally, we have compared with the
photoemission studies to find that the new band around the Fermi
energy of the $A_1$$\soc$ phase can be explaind by the fact that
a energy level intrudes extremely largely into the gap of the
neutral system by the electron doping.  This is another indication
that the interaction effects among electrons play a crucial role
in doped $\soc$-polymers.

\pagebreak
\begin{flushleft}
{\bf REFERENCES}
\end{flushleft}

\noindent
$^1$ O. Chauvet, G. Oszl\`{a}nyi, L. Forr\'{o}, P. W. Stephens,
M. Tegze, G. Faigel, and A. J\`{a}nossy,
Phys. Rev. Lett. {\bf 72}, 2721 (1994).\\
$^2$ P. W. Stephens, G. Bortel, G. Faigel, M. Tegze,
A. J\`{a}nossy, S. Pekker, G. Oszlanyi, and L. Forr\'{o},
Nature {\bf 370}, 636 (1994).\\
$^3$ S. Pekker, L. Forr\'{o}, L. Mihaly, and A. J\`{a}nossy,
Solid State Commun. {\bf 90}, 349 (1994).\\
$^4$ S. Pekker, A. J\`{a}nossy, L. Mihaly, O. Chauvet,
M. Carrard, and L. Forr\'{o}, Science {\bf 265}, 1077 (1994).\\
$^5$ Y. J. Uemura, K. Kojima, G. M. Luke, W. D. Wu,
G. Oszlanyi, O. Chauvet, and L. Forr\'{o}, Phys. Rev. B
{\bf 52}, R6991 (1995).\\
$^6$ K. Tanaka, Y. Matsuura, Y. Oshima, T. Yamabe,
Y. Asai, and M. Tokumoto, Solid State Commun.
{\bf 93}, 163 (1995).\\
$^7$ P. R. Surj\'{a}n and K. N\'{e}meth, Solid State Commun.
{\bf 92}, 407 (1994).\\
$^8$ W. P. Su, J. R. Schrieffer, and A. J. Heeger,
Phys. Rev. B {\bf 22}, 2099 (1980).\\
$^9$ K. Harigaya, Phys. Rev. B {\bf 52}, 7968 (1995).\\
$^{10}$ K. Harigaya, Chem. Phys. Lett. {\bf 242}, 585 (1995).\\
$^{11}$ K. Tanaka, Y. Matsuura, Y. Oshima, T. Yamabe,
H. Kobayashi, and Y. Asai, Chem. Phys. Lett. {\bf 241}, 149 (1995).\\
$^{12}$ K. Harigaya and S. Abe, Phys. Rev. B {\bf 49}, 16746 (1995).\\
$^{13}$ G. P. Lopinski, M. G. Mitch, J. R. Fox, and J. S. Lannin,
Phys. Rev. B {\bf 50}, 16098 (1994).\\
$^{14}$ K. Harigaya, Phys. Rev. B {\bf 45}, 13676 (1992).\\

\pagebreak

\begin{flushleft}
{\bf FIGURE CAPTIONS}
\end{flushleft}

\mbox{}

\noindent
Fig. 1.  The crystal structures of the one-dimensional
$\soc$-polymer.  The carbon sites which constitute the
four membered rings are named with numbers.
The bond lengths, used in the calculations, are also shown.

\mbox{}

\noindent
Fig. 2.  The magnitude of the magnetization per one $\soc$
shown by changing $a$ and $U$.  We use $V = U/2$ here.
The closed and open squares are for $a= 0.1$ and $0.2$,
respectively.  The closed and open circles are for
$a=0.3$ and $0.4$.

\mbox{}

\noindent
Fig. 3.  The phase diagram shown against $a$ and $U$.
The region with ``M" is the metallic phase, and
the region with ``AF" is the antiferromagnetic phase.

\mbox{}

\noindent
Fig. 4.  The band structures of the neutral $\soc$-polymer
of the case $(a,U,V)=(0.3, 2.4t, 1.2t)$.  The unit cell
consists of one $\soc$.   The length of the unit cell is taken
as unity.  The highest fully occupied band is named as ``HOMO",
and the lowest empty band as ``LUMO".

\mbox{}

\noindent
Fig. 5.  The band structures of the $\soc$-polymer doped with
one electron per one $\soc$ of the case $(a,U,V)=(0.3, 2.4t, 1.2t)$.
The unit cell consists of two $\soc$ molecules.   The length of
the unit cell is taken as unity.  The highest fully occupied band
is named as ``HOMO", and the lowest empty band as ``LUMO".

\end{document}